\documentclass[prx,onecolumn,preprintnumbers,superscriptaddress,nofootinbib,longbibliography, 11pt]{revtex4-2} 

\usepackage{graphicx}
\usepackage{bm}
\usepackage{amsmath}
\usepackage{amssymb}
\usepackage{amsfonts}
\usepackage{slashed}
\usepackage{latexsym,epsfig,bbm}
\usepackage{algorithm}
\usepackage{algorithmic}

\newcommand{\bra}[1]{\langle{#1}|}
\newcommand{\ket}[1]{|{#1}\rangle}

\newcommand{\braket}[2]{\langle{#1}|{#2}\rangle}
\newcommand{\bopk}[3]{\langle{#1}|{#2}|{#3}\rangle}

\newcommand{\figref}[1]{Fig.~\ref{#1}}

\usepackage{color}
\definecolor{blue}{rgb}{0,0.2,1}

\definecolor{red}{rgb}{0.9,0,0}

\newcommand{\past}[1]{\overleftarrow{#1}}
\newcommand{\fut}[1]{\overrightarrow{#1}}

\newcommand{\tpasto}{t_{\overleftarrow{0}}}

\newcommand{\tfuto}{t_{\overrightarrow{0}}}

\newcommand{\tpast}{\overleftarrow{t}}

\newcommand{\tfut}{\overrightarrow{t}}

\begin{document}

\title{Embedding memory-efficient stochastic simulators as quantum trajectories}

\author{Thomas J.~Elliott}
\email{physics@tjelliott.net}
\affiliation{Department of Physics \& Astronomy, University of Manchester, Manchester M13 9PL, United Kingdom}
\affiliation{Department of Mathematics, University of Manchester, Manchester M13 9PL, United Kingdom}
\affiliation{Department of Mathematics, Imperial College London, London SW7 2AZ, United Kingdom}
\affiliation{Complexity Institute, Nanyang Technological University, Singapore 637335}
\author{Mile Gu}
\email{mgu@quantumcomplexity.org}
\affiliation{Nanyang Quantum Hub, School of Physical and Mathematical Sciences, Nanyang Technological University, Singapore 637371}
\affiliation{Complexity Institute, Nanyang Technological University, Singapore 637335}
\affiliation{Centre for Quantum Technologies, National University of Singapore, 3 Science Drive 2, Singapore 117543}
\affiliation{MajuLab, CNRS-UNS-NUS-NTU International Joint Research Unit, UMI 3654, Singapore 117543}

\date{\today}

\begin{abstract}
By exploiting the complexity intrinsic to quantum dynamics, quantum technologies promise a whole host of computational advantages. One such advantage lies in the field of stochastic modelling, where it has been shown that quantum stochastic simulators can operate with a lower memory overhead than their best classical counterparts. This advantage is particularly pronounced for continuous-time stochastic processes; however, the corresponding quantum stochastic simulators heretofore prescribed operate only on a quasi-continuous-time basis, and suffer an ever-increasing circuit complexity with increasing temporal resolution. Here, by establishing a correspondence with quantum trajectories -- a method for modelling open quantum systems -- we show how truly continuous-time quantum stochastic simulators can be embedded in such open quantum systems, bridging this gap and obviating previous constraints. We further show how such an embedding can be made for discrete-time stochastic processes, which manifest as jump-only trajectories, and discuss how viewing the correspondence in the reverse direction provides new means of studying structural complexity in quantum systems themselves.
\end{abstract}
\maketitle 

\section{Introduction}

One of the hallmark features of quantum systems is that they appear complex to our classical intuitions. Indeed, the simulation of many-body quantum systems with classical computers is a challenge that grows exponentially with each additional particle. This led to one of the first proposed applications of quantum computers -- simulation of quantum systems -- leveraging their intrinsically quantum nature to escape this cursed scaling of complexity~\cite{feynman1982simulating}.

We can extract further utility from this innate complexity, by employing quantum technologies to perform other complex computations~\cite{nielsen2000quantum, preskill2018quantum}. A growing body of research has explored the application of quantum technologies in the simulation of (classical) stochastic dynamics, finding that such quantum stochastic simulators can operate with a lower memory cost and smaller thermodynamical footprint than possible with any classical simulator~\cite{gu2012quantum, mahoney2016occam, liu2019optimal, loomis2020thermal, elliott2021memory}. These quantum advantages have been theoretically proven to exhibit favourable scaling~\cite{aghamohammadi2017extreme, garner2017provably, thompson2018causal}, especially when simulating continuous-time stochastic processes~\cite{elliott2018superior, elliott2019memory, elliott2020extreme, elliott2021quantum, wu2023implementing}. Thus far however, explicit proposals for the construction of such models are based on discrete-time quantum evolutions, approaching only a \emph{quasi}-continuous evolution in the limit of performing infinitesimal quantum gates in rapid succession. This presents a significant practical barrier to the demonstration of the scalability of the quantum memory advantage. 

Here, we remove this barrier with a proposal for a truly continuous-time quantum stochastic simulator. Our proposal consists of embedding continuous-time quantum simulators within the evolution of a naturally continuous-time open quantum system. By mapping the infinitesimal Kraus operators of the quasi-continuous quantum simulators into an appropriate Hamiltonian and set of dissipators for the open system we are able to specify a continuous-time simulator, with the state of the system acting as the memory, and the outputs obtained by monitoring the dissipation channels. That is, the statistics of the trajectories of the open system evolution correspond to the statistics of the simulated process.

We begin with a recapitulation of the relevant background on stochastic processes, (quantum) models of their evolution, and the open system trajectory formalism in Section \ref{secframework}. We then establish the mapping by which continuous-time quantum simulators can be embedded within open quantum system trajectories in Section \ref{secct}, followed by an analogous mapping for discrete-time simulators in Section \ref{secdt}. We briefly outline how this embedding may also provide an interesting lens through which the complexity of \emph{quantum} systems and processes can be characterised in Section \ref{secrev}, and conclude in Section \ref{secdis}.

\section{Framework}
\label{secframework}
\subsection{Stochastic Processes and Models}

A continuous-time, discrete-event stochastic process~\cite{marzen2017structure} consists of a probabilistic series of observable events $x_n\in\mathcal{X}$, where the subscript $n$ denotes the event number. The time between the $(n-1)$th and $n$th events is denoted by $t_n\in\mathbb{R}^+$ (which is itself also typically a stochastic variable); for shorthand we denote $\bm{x}_n:=(x_n,t_n)$, and $x_{l:m}:=x_lx_{l+1}\ldots x_{m-1}$ represents a string of consecutive events. The dynamics of such processes are typically governed by an underlying hidden system, and the observed events are described by a collective distribution $P(\ldots,\bm{X}_{n-1},\bm{X}_n, \bm{X}_{n+1},\ldots)$; we use upper and lower case to distinguish random variables from their corresponding variates. Here we consider stationary (i.e., time-invariant) stochastic processes, such that $n\in\mathbb{Z}$ and $P(\bm{X}_{0:L})=P(\bm{X}_{m:m+L})\forall m,L\in\mathbb{Z}$.

We can divide the process into a past and future, describing the events that have occured thus far, and those yet to occur respectively. That is, the past $\past{\bm{x}}:=\lim_{L\to\infty}\bm{x}_{-L:0}(\emptyset,\tpasto)$, where without loss of generality we have taken $x_0$ to be the next event to occur, $\tpasto$ is the time since the last event, and $\emptyset$ represents that the $0$th event is yet to occur. Similarly, the future $\fut{\bm{x}}:=\lim_{L\to\infty}(x_0,\tfuto)\bm{x}_{1:L}$, where $\tfuto$ is the time until the next event, such that $t_0=\tpasto+\tfuto$.

A causal model of a stochastic process~\cite{crutchfield2012between} uses information about the past of the process to produce a series of future events commensurate with the statistics of the process. An (exact) model is able to simulate these statistics perfectly, such that given any past $\past{\bm{x}}$, the model produces futures $\fut{\bm{X}}$ with the same probabilities as the process' conditional distribution $P(\fut{\bm{X}}|\past{\bm{x}})$. To do this, the model encodes the relevant information from the past into a memory; this is achieved with an encoding function $f:\past{\mathcal{\bm{X}}}\to\mathcal{M}$, where $\rho_m\in\mathcal{M}$ are the states of the memory. The model also needs a means of evolution that produces the outputs and updates the memory, i.e., a dynamic $\Lambda:\mathcal{M}\to\mathcal{M},\emptyset\cup\mathcal{X}$ acting continuously. A key metric of performance for the model is the amount of memory it requires: two such metrics are the statistical and topological memories~\cite{crutchfield1989inferring}, respectively defined as
\begin{align}
\label{eqcosts}
D_f&:=\log_2[\mathrm{rank}(\rho)].\nonumber\\
C_f&:=-\mathrm{Tr}(\rho\log_2[\rho]),
\end{align}
capturing the (log of) dimensions required by the memory, and the amount of information it must store. Here, $\rho:=\sum_mP(m)\rho_m$ is the steady-state of the memory, with $P(m):=\sum_{\past{\bm{x}}:f(\past{\bm{x}})=\rho_m}P(\past{\bm{x}})$.

The provably memory minimal classical model (according to both measures) can be systematically found using the tools of computational mechanics~\cite{crutchfield1989inferring, shalizi2001computational, crutchfield2012between}, a branch of complexity science. The causal states of a process are defined according to an equivalence relation $\sim_\varepsilon$ clustering together pasts iff they have identical future statistics, i.e., $\past{\bm{x}}\sim_\varepsilon\past{\bm{x}}'\Leftrightarrow P(\fut{\bm{X}}|\past{\bm{x}})=P(\fut{\bm{X}}|\past{\bm{x}}')$. The causal state encoding function $f_\varepsilon$ then maps pasts to the same (classical) memory state iff they belong to the same causal state. The statistics of the process then define the transition dynamic between these states, and the corresponding model is referred to as the $\varepsilon$-machine of the process. For typical continuous-time processes, these measures (labelled $D_\mu$ and $C_\mu$) are both infinite in the truly-continuous limit~\cite{marzen2017informational, marzen2017structure, elliott2018superior}, requiring coarse-grained discrete-time approximations for finite memory realisations~\cite{marzen2015informational, marzen2015time}.

\begin{figure}
\includegraphics[width=\linewidth]{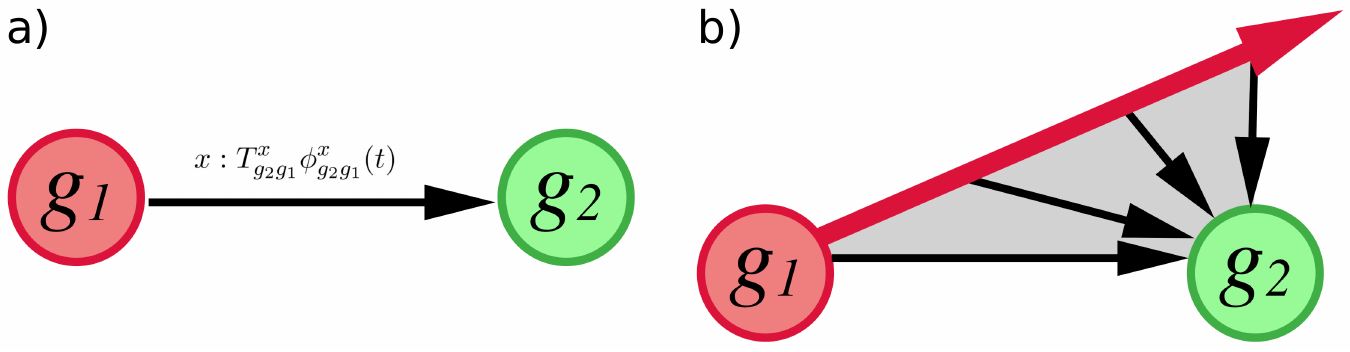}
\caption{(a) Hidden semi-Markov model representation of transitions in a stochastic process. The notation $x:T_{g_2g_1}^x\phi_{g_2g_1}^x(t)$ denotes the probability that a process that is currently in mode $g_1$ immediately after an event will next display event $x$ after a time $t$ and transition into mode $g_2$ is given by $T_{g_2g_1}^x\phi_{g_2g_1}^x(t)$. (b) Hidden Markov model representation of the same transition, where time since last event is tracked by a continuum of states represented by the red arrowed line.}
\label{fighsmmhmm}
\end{figure}

Under weak assumptions on the process~\cite{marzen2017structure}, the causal states can be labelled by a pair $(g,t)$, where $g\in\mathcal{G}$ is referred to as a mode, and $t$ is the time since last event. Given a model in state $(g,t)$, in the next infinitesimal interval $dt$ the model will evolve to $(g,t+dt)$ if no event occurs, or proceed to $(g',0)$ if an event does take place, where the new mode $g'$ is a deterministic function of the previous mode $g$ and event $x$. To each mode $g$ we can assign a series of distributions $T_{g'g}^x\phi_{g'g}^x(t)$ describing the probability that a model resides for a dwell time $t$ in mode $g$ (i.e., the inter-event duration) before event $x$ occurs and a transition to mode $g'$ occurs. This can be represented by a hidden semi-Markov model (HSMM)~\cite{marzen2017structure}, and further unwravelled as a (continuous- or discrete-state) hidden Markov model (HMM)~\cite{rabiner1986introduction}, as depicted in \figref{fighsmmhmm}.

\subsection{Quantum Stochastic Simulators}
\label{secqss}

\begin{figure}
\includegraphics[width=0.8\linewidth]{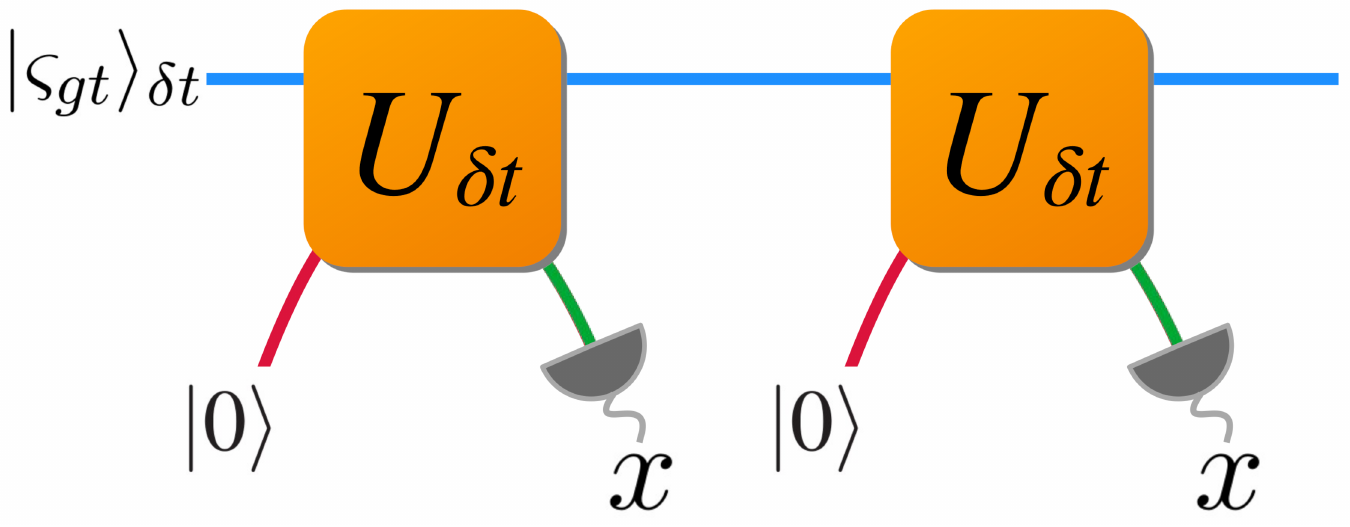}
\caption{Quantum circuit representation of a quasi-continuous quantum simulator, showing two timesteps of evolution. At each timestep, the current memory state $\ket{\varsigma_{gt}}$ (blue wire) undergoes a joint interaction $U_{\delta t}$ (orange box) together with an ancilla (red wire) initially in blank state $\ket{0}$, to produce an updated memory state and an output ancilla (green wire) that produces the output for that timestep following measurement in the computational basis. The memory state is then fed forwards into the evolution for the next timestep, together with a fresh blank ancilla.}
\label{figcircuit}
\end{figure}

While $\varepsilon$-machines are minimal amongst classical models, quantum models can do better~\cite{gu2012quantum, elliott2019memory}. Such quantum models use an encoding function $f_q$ that maps pasts to quantum (i.e., non-mutually orthogonal) memory states~\cite{elliott2020extreme}. The current state-of-the-art constructions~\cite{liu2019optimal, elliott2021quantum} follow $f_\varepsilon$ in clustering pasts according to the causal states, but now with quantum memory states $\{\ket{\varsigma_{gt}}_{\delta t}\}$ in their place, with the subscript $\delta t$ indicating the implicit dependence on the coarse-graining into finite-sized timesteps. The quantum memory states are defined implicitly according to a quasi-continuous evolution operator $U_{\delta t}$:
\begin{equation}
\label{eqexactct}
U_{\delta t}\ket{\varsigma_{gt}}_{\delta t}\ket{0}=\sqrt{\frac{\Phi_g(t+\delta t)}{\Phi_g(t)}}\ket{\varsigma_{gt+\delta t}}_{\delta t}\ket{0}
+\sum_{xg'}\sqrt{\frac{\int_t^{t+\delta t}T_{g'g}^x\phi_{g'g}^x(t')dt'}{\Phi_g(t)}}\ket{\varsigma_{g'0}}_{\delta t}{\ket{x}},
\end{equation}
where the (modal) survival probability $\Phi_g(t):=\sum_{xg'}\int_{t}^\infty T_{g'g}^x\phi_{g'g}^x(t')dt'$ represents the probability that the system will remain in mode $g$ for at least time $t$. The first system contains the memory, while the second is an ancilla that probes the memory to produce the event statistics; 0 is used as a proxy for no event $\emptyset$. Each application of $U_{\delta t}$ representes one timestep of evolution, with a fresh ancilla introduced for each such timestep. See \figref{figcircuit} for a schematical quantum circuit depicting this evolution.

The corresponding memory measures ($D_q$ and $C_q$, taken in the limit $\delta t\to0$) satisfy $D_q\leq D_\mu$ and $C_q\leq C_\mu$, with the inequalities strict whenever the quantum memory states are linearly dependent or have non-zero overlap respectively~\cite{liu2019optimal}. Typically, one finds that $C_q$ is finite (in contrast to the classical divergence), while strict advantages of $D_q$ are known only for specific families of processes~\cite{thompson2018causal, ghafari2019dimensional, elliott2020extreme, wu2023implementing} -- though recent work has developed methods for drastically reducing the memory dimension required of near-exact quantum models to simulate continuous-time stochastic processes~\cite{elliott2021quantum}. 

\subsection{Quantum Trajectory Formalism}

While an idealised, closed quantum system evolves according to unitary operators driven by a Hermitian Hamiltonian, in reality this picture typically breaks down. Open quantum systems are coupled to and exchange information with their environment. Under the approximation that the environment is left largely unchanged by its coupling to the system, the evolution of the system can be described by repeated application of a quantum channel -- or in the continuous-time limit, a Lindbladian Markovian quantum master equation~\cite{breuer2002theory}:
\begin{equation}
\label{eqmast}
\frac{d\rho}{dt}=-i[H,\rho]-\frac{1}{2}\sum_j\gamma_j (c_j^\dagger c_j\rho+\rho c_j^\dagger c_j-2c_j\rho c_j^\dagger),
\end{equation}
where $\rho$ is the state of the system, $H$ is its `natural' closed system Hamiltonian, $\gamma_j$ are the strengths (i.e., rates) of a series of dissipative channels and $c_j$ the associated action (`jump') on the state following said dissipation. $[A,B]:=AB-BA$ represents the standard commutator and we have set $\hbar=1$ for convenience. Often, as we shall do here, the dissipation operators are rescaled and normalised according to their rates to give the jump operators $J_j:=\sqrt{\gamma_j}c_j$. Moreover, an effective Hamiltonian can be prescribed, describing the (non-Hermitian) evolution of the system in the event that no dissipation occurs: $H_\mathrm{eff}:=H-(i/2)\sum_jJ^\dagger_jJ_j$, with the norm of the state corresponding to the probability that no such dissipation would have occured within the evolution time. 

Nevertheless, solving the quantum master equation is often computationally taxing, especially since it requires one to propagate the full density matrix of the system. A powerful alternative approach is the \emph{quantum trajectory} formalism~\cite{carmichael1993quantum, wiseman1996quantum, wiseman2009quantum, daley2014quantum}, also referred to as the Monte Carlo wavefunction method~\cite{molmer1993monte} or the quantum jump formalism~\cite{plenio1998quantum}. The premise of this approach is to record all jumps, and conditionally update our description of the state accordingly. Supposing that we do indeed have such a record of all jumps and when they occur (a `trajectory'), then given an initial pure state of the system our description of the system remains pure at all times.

Note that the decomposition of a given quantum master equation into a set of dissipative channels is not unique. Indeed, given a set of jump operators $\{J_j\}$, the same evolution can be obtained from a master equation with jump operators $\{J_j'\}$ resulting from a unitary reshuffling of the labels of the original jump operators, i.e., $J_k'=\sum_ju_{kj}J_j$ for some unitary matrix $u$. Thus, the unwravelling of a quantum master equation into a set of trajectories is not unique, and depends on the choice of jump operators. In practical terms, this corresponds to the choice in how the dissipation is monitored (i.e., measured).

There are two stages to the evolution of a system on a particular trajectory. Between jump events the system evolves according to the effective Hamiltonian, i.e., 
\begin{equation}
\ket{\psi(t)}=U_{\mathrm{eff}}(t-t')\ket{\psi(t')},
\end{equation}
where $U_{\mathrm{eff}(t)}:=\exp(iH_{\mathrm{eff}}t)$. Note that due to the non-Hermiticity of $H_{\mathrm{eff}}$, $U_{\mathrm{eff}}$ is non-unitary and hence does not preserve normalisation of the state; thus, one must appropriately rescale the state normalisation at the end of the evolution. The second stage is the effect of the jumps. Upon jump $j$, (the conditional description of) the state undergoes the instantaneous transformation 
\begin{equation}
\ket{\psi(t)}\to J_j\ket{\psi(t)}.
\end{equation}
This also does not preserve the normalisation of the state, and so also requires an appropriate rescaling. Such rescalings notwithstanding, the norms of the non-normalised states carry physical significance. For evolution under the effective Hamiltonian, the norm describes the probability of the system surviving for that length of time without undergoing a decay event. Meanwhile, the norm of the post-jump state describes the instantaneous probability per unit time of the specified jump event occuring. Given a set of jump operators $\{J\}$ and a natural Hamiltonian $H$, the trajectory uniquely specifies a conditional evolution of the system. A weighted average over all possible trajectories will recover the stochastic ensemble evolution of the density matrix as described by the master equation.

Compared to master equations, simulation of a quantum trajectory is comparatively more efficient. By sampling over many such trajectories, one can estimate properties of the open system, such as expectation values and correlations. A standard approach to this sampling is as follows. Beginning with an initial state $\ket{\psi(0)}$, generate a random number $r\in[0,1]$, and determine the time $t$ such that $\bopk{\psi(0)}{U_{\mathrm{eff}}^\dagger(t)U_{\mathrm{eff}}(t)}{\psi(0)}=r$; this specifies that a jump event occurs at time $t$. To determine which jump occurs, randomly choose one of the $j$ weighted according to $\bopk{\psi(t)}{J^\dagger_j J_j}{\psi(t)}$. Repeat the above steps starting from the (rescaled to unit norm) post-jump state $J_j\ket{\psi(t)}$, until the maximum time of the simulation is reached. This generates a trajectory with the appropriate weighting. By generating many such trajectories the sampling can be performed.

\section{Embedding continuous-time quantum stochastic simulators}
\label{secct}

The circuit-based picture of quantum stochastic simulators described in Sec.~\ref{secqss} allows us to consistently define memory states $\ket{\varsigma_{gt}}$ for all $g\in\mathcal{G}, t\in\mathbb{R}^+$ (i.e., for all possible continuum causal states). However, the evolution is implicitly only quasi-continuous, discretised into timesteps of size $\delta t$. While this can in principle be refined arbitrarily, a fresh probe ancilla is required at each timestep, as well as an ever-increasing number of gates. Specifically, to simulate the statistics up to some fixed time the number of ancillas and the number of calls to $U_{\delta t}$ must scale at least as fast as inversely proportional to the size of the timesteps, no matter how efficiently $U_{\delta t}$ itself can be implemented. This is because we are required to sequentially produce the output statistics that correspond to measurement of the ancillas; thus, while there are powerful techniques for reducing the complexity of circuit-based simulation of Lindbladian dynamics~\cite{cleve2016efficient, ramusat2021quantum}, as they do not produce the same observable behaviour on their ancillary systems, we cannot make use of them here. We are effectively studying the dual of the problem here, in that our objective is to determine how to implement an open system that gives rise to the desired statistics we wish to simulate, rather than simulation of any particular open system in itself.

We will now overcome this issue by embedding the model into the dynamics of an open quantum system, such that the statistics of the process are mapped to quantum trajectories. Indeed, the jump events of a quantum trajectory are themselves a continuous-time stochastic process. Here we show how the jump operators $\{J_j\}$ and effective Hamiltonian $H_{\mathrm{eff}}$ (and thus natural Hamiltonian $H$) can be designed such that this process corresponds to that which we desire to simulate. Further, the state of the open system at any point in the trajectory is identical (up to unitary symmetry) to the analogous memory state of the quantum model specified Eq.~\eqref{eqexactct}. Viewing the state of the open system as a memory\footnote{Indeed, the Markovian nature of the master equation unwravelled by a quantum trajectory guarantees that any memory in the dynamics must be contained within the system.}, the system thus forms a quantum stochastic simulator of the process with the same memory costs Eq.~\eqref{eqcosts} as the quantum model Eq.~\eqref{eqexactct}, but with a truly continuous-time evolution.

To make this mapping, we must first assign the quantum memory states and evolution of the quasi-continuous model. The overlaps of the quantum memory states can be obtained from Eq.~\eqref{eqexactct}, using that $\braket{\varsigma_{gt}}{\varsigma_{g't'}}_{\delta t}=\bra{\varsigma_{gt}}\bopk{0}{U^\dagger_{\delta t}U_{\delta t}}{\varsigma_{g't'}}_{\delta t}\ket{0}$. These can then be assigned in terms of an arbitrary basis using a reverse Gram-Schmidt procedure. The columns of $U_{\delta t}$ prescribed by the model definition can then be expressed in this basis, and the remainder of the columns can be assigned arbitrarily, provided all columns are mutually orthogonal. See e.g., Refs.~\cite{binder2018practical, liu2019optimal, elliott2021quantum} for further details.

From this unitary operator, we are able to designate a set of Kraus operators corresponding to each of the possible events $K^x_{\delta t}:=(\mathbb{I}\otimes\bra{x})U_{\delta t}(\mathbb{I}\otimes\ket{0})$, capturing the effective evolution of the memory conditioned on event $x$ occuring ($\mathbb{I}$ is the identity matrix, here acting on the memory subsystem). Similarly, we can designate $K^0_{\delta t}:=(\mathbb{I}\otimes\bra{0})U_{\delta t}(\mathbb{I}\otimes\ket{0})$ for the non-event evolution.

For the majority of the timesteps, the system will not exhibit an event, and will instead undergo the non-event evolution. Indeed, it can be seen that $\ket{\varsigma_{gt+\delta t}}\propto K_{\delta t}^0\ket{\varsigma_{gt}}$, and $\ket{\varsigma_{gn\delta t}}\propto (K^0_{\delta t})^n\ket{\varsigma_{g0}}$. This parallels the action of the non-Hermitian evolution under $H_{\mathrm{eff}}$ in a quantum trajectory, accounting for the gradual shift in belief of the memory/system state conditioned on the lack of events occuring. Conversely, the Kraus operators $K^x_{\delta t}$ corresponding to the events occur with much lower probability on each timestep, and give rise to much sharper transitions in the system state, abruptly placing the memory in the start state of a new mode. This resembles the action of the jump operators in a quantum trajectory. Note however that we require a specific choice of jump operators to obtain the desired model, and thus lose the freedom in how we unwravel the associated master equation. That is, we must monitor the dissipation from the open memory system in a particular manner to manifest the target statistics.

Let us cast the non-event evolution in terms of an evolution under a non-Hermitian Hamiltonian $H^{\mathrm{NH}}_{\delta t}$, i.e., $K^0_{\delta t}=\exp(-iH^{\mathrm{NH}}_{\delta t}\delta t)$. For a consistent trajectory, we require that the non-Hermitian Hamiltonian is consistent for all timestep sizes $\delta t$, whereupon we can replace it by $H_{\mathrm{eff}}$. This is possible if the infinitesimal evolution can be generated at all times, for all modes; by expanding $\exp(-iH_{\mathrm{eff}}\delta t)\approx \mathbb{I}-iH_{\mathrm{eff}}\delta t$ for small $\delta t$, we thence require $\mathbb{I}-\braket{\varsigma_{gt}}{\varsigma_{gt+\delta t}}\propto\delta t \forall  g, t$. In Appendix A, we show that this is satisfied when all modal distributions $\phi_{g'g}^x(t)$ are everywhere finite and almost-everywhere continuous; these are natural conditions to expect of a physically-reasonable continuous-time stochastic process. Then, we are able to associate the non-Hermitian evolution of the trajectory with
\begin{equation}
\label{eqheffembed}
H_{\mathrm{eff}}=\lim_{\delta t\to0} \frac{\ln K^0_{\delta t}}{-i\delta t}\approx \lim_{\delta t\to0}\frac{\mathbb{I}-K^0_{\delta t}}{i\delta t}.
\end{equation}

It is then comparatively straightforward to deduce the relationship between the the Kraus operators corresponding to events and the jump operators of the associated trajectory: for sufficiently small timesteps $\delta t$ the Kraus operators \emph{are} the rate-normalised jump operators, scaled by $\sqrt{\delta t}$. This can be seen by first noting that the probability of symbol $x$ being emitted in the next interval $\delta t$ given current mode $g$ and time since last event $t$ is given by $P_{\delta t}(x|g,t):=\sum_{g'}\int_t^{t+\delta t}T^x_{g'g}\phi_{g'g}^x(t')dt'/\Phi_g(t)$. Then, using that only one term in the sum is non-zero (since the subsequent mode is a deterministic function of $g$ and $x$), we have that $P_{\delta t}(x|g,t)=\bopk{\varsigma_{gt}}{{K^x_{\delta t}}^\dagger K^x_{\delta t}}{\varsigma_{gt}}$. Comparing this to the probabilities associated with the jump operators of a trajectory (namely, that the probability of jump $x$ occuring in the next infinitesimal interval $dt$ given current state $\ket{\psi}$ is $\bopk{\psi}{J_x^\dagger J_x}{\psi}dt$), it follows that
\begin{equation}
\label{eqjump}
J_x=\lim_{\delta t\to0}\frac{K^x_{\delta t}}{\sqrt{\delta t}}.
\end{equation}
It can readily be seen that this limit exists and is well-defined under the conditions placed on the $\phi_{g'g}^x(t)$ for the non-Hermitian evolution to also be well-defined, namely, that they are everywhere finite and almost-everywhere continuous.

\begin{figure}
\includegraphics[width=\linewidth]{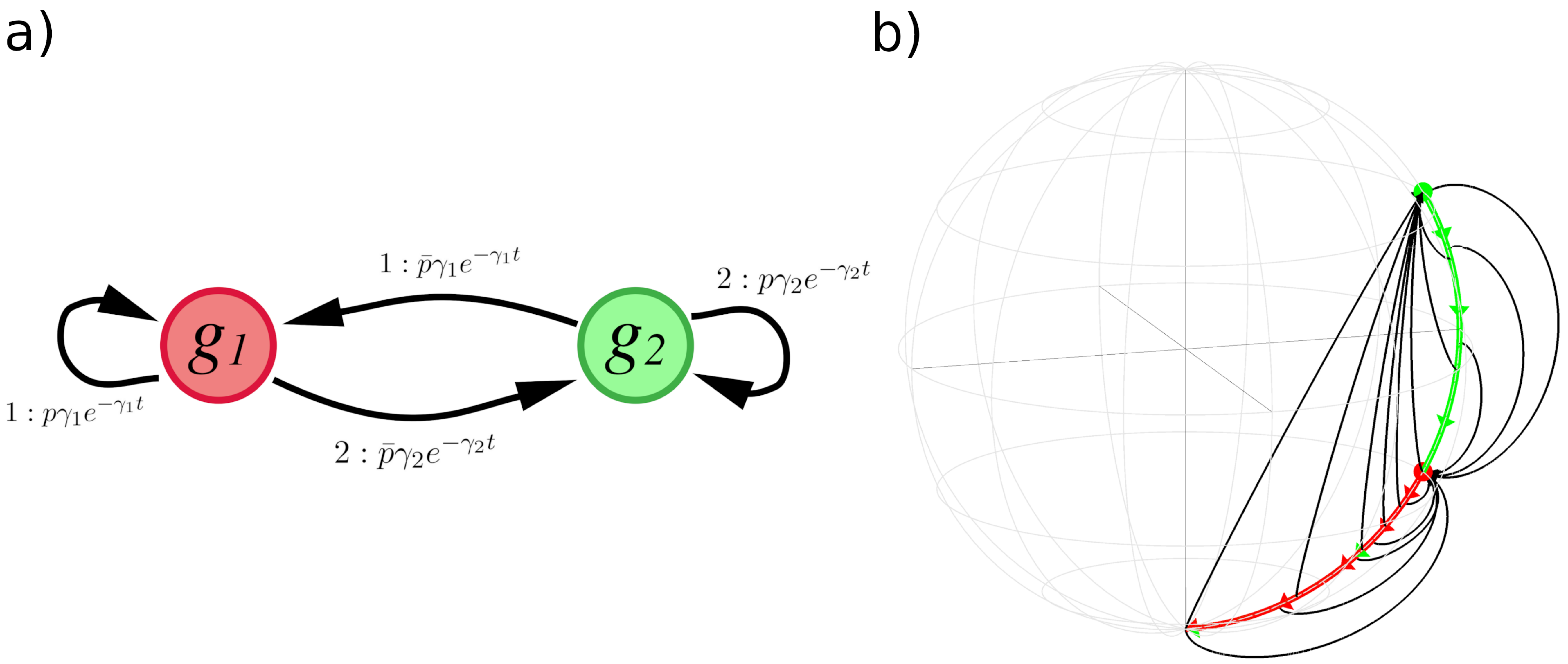}
\caption{(a) HSMM representation of the example two channel decay process described in the main text. The process has two modes that assign different weights to each channel, heralded by the symbol emitted in the previous decay. (b) Bloch sphere representation of the quantum memory states, with the thick coloured arrows depicting the continuous space of quantum memory states, and the thinner black arrows the transitions upon decay events. Plot shown for $p=0.25$, $\gamma_1\neq\gamma_2$; the specific values of the decay rates are otherwise arbitrary as they do not impact the state space, only the rate at which it is traversed.}
\label{fighsmmex}
\end{figure}

As an example, let us consider the following process. A system undergoes a series of decays from a pair of decay channels, with associated rates $\gamma_1$ and $\gamma_2$ respectively. Each decay is heralded by an event signifying which channel the decay came from. The choice of channel is probabilistically assigned, and hidden, such that a causal model must track a belief in the likelihood of which channel was chosen, based on the time since last event. This choice of channel also varies based on the last event, such that if the last decay was from channel 1, the weightings are $p$ for channel 1 and $\bar{p}:=1-p$ for channel 2, and reversed if the last decay was from channel 2. Thus, there is an alphabet $\mathcal{X}=\{1,2\}$, and two modes $\mathcal{G}=\{g_1,g_2\}$. The HSMM representation of the process is given in \figref{fighsmmex}(a), with $T^{x}_{g_{x'}g_{x''}}$ taking value $p$ if $x=x'=x''$, $\bar{p}$ if $x=x'\neq x''$, and zero otherwise, and $\phi^x_{g_{x'}g_{x''}}(t)=\gamma_x\exp(-\gamma_xt)$. This can be seen as a generalisation of the dual Poisson process, previously used to demonstrate extreme dimensional memory advantages of quantum models~\cite{elliott2020extreme}.

In Appendix B we show how this process can be exactly modelled causally by a quantum system with a single qubit memory for all values of $\gamma_1$, $\gamma_2$, and $p$. Meanwhile, we also show that the minimal exact classical causal model requires an infinite memory dimension. We further give expressions for the Kraus operators of our quantum model. Correspondingly, we obtain that the associated trajectory is described by:
\begin{equation}
H_{\mathrm{eff}}=\begin{pmatrix}-\frac{i\gamma_1}{2} & 0\\ 0 & -\frac{i\gamma_2}{2}\end{pmatrix},\quad\quad\quad\quad J_1=\begin{pmatrix}\sqrt{\gamma_1p} & 0\\ \sqrt{\gamma_1\bar{p}} & 0\end{pmatrix},\quad\quad\quad\quad J_2=\begin{pmatrix}0 & \sqrt{\gamma_2\bar{p}}\\0 & \sqrt{\gamma_2p}\end{pmatrix}.
\end{equation}
We illustrate this in \figref{fighsmmex}(b) for a representative set of parameters, showing the possible paths of the trajectory through the memory state space.

\section{Embedding discrete-time quantum stochastic simulators}
\label{secdt}

A discrete-time, discrete-event stochastic process with alphabet $\mathcal{X}$ is specified by the distribution $P(\ldots,X_{n-1},X_n,X_{n+1},\ldots)$. This describes symbol-only dynamics, where an event occurs at each interval, in which case $\mathcal{X}$ is the set of possible events. Similarly, it also descibes coarse-grainings of the continuous-time processes detailed above, where the time variable is discretised into finite timesteps of $\Delta t$. In this latter case, $\mathcal{X}$ is the union of the set of possible events and a null event $\emptyset$, that denotes intervals when no event occured. 

As with continuous-time processes, the causal states are defined through an equivalence relation on the set of pasts, i.e., $\past{x}\sim_\varepsilon\past{x}'\Leftrightarrow P(\fut{X}|\past{x})=P(\fut{X}|\past{x}')$, where the past now consists only of value of $\past{X}$ at each past interval. These can again be used to specify an encoding $f_\varepsilon$ that prescribes the provably memory minimal classical model ($\varepsilon$-machine) of the process. However, as this now encodes into a discrete state space, these models are now representable by hidden Markov models. 

We can similarly construct quantum models of such processes with sub-classical memory costs, using the following prescription:
\begin{equation}
\label{eqdisc}
U\ket{\sigma_s}\ket{0}=\sum_{x,s'}\sqrt{T^x_{s's}}e^{i\varphi_{xs}}\ket{\sigma_{s'}}\ket{x},
\end{equation}
where $\{\ket{\sigma_s}\}$ are the quantum memory states, the second subspace is measured to give the measurement outcome, $\{\varphi_{xs}\}$ are an arbitrary set of phases,\footnote{Note that the choice of phases does impact the memory costs Eq.~\eqref{eqcosts}. Primarily, they have been used to engineer linear dependencies between memory states in order to reduce $D_q$~\cite{liu2019optimal}.} $T^x_{s's}$ is the probability that the event $x$ occurs given we started in state $s$, and the updated state $s'$ is a deterministic function of $s$ and $x$.

By taking the interval between events to be a stochastic variable, a discrete-time stochastic process can be instantiated within a continuous-time stochastic process, with the events of the latter process corresponding to the events (or null events) of the original discrete time process. That is, each event in the continuous-time process corresponds to one interval in the discrete-time process. Moreover, when the stochastic variable governing the time interval between events is memoryless (i.e., it takes the form of an exponential decay), then this does not require any additional memory to model beyond that of original discrete-time process.

Let this decay rate be $\gamma$, such that the probability of the next event occuring within time $t$ from the present is $1-\exp(-\gamma t)$. Then, we can represent the quasi-continuous evolution in the form of Eq.~\eqref{eqexactct}:
\begin{equation}
\label{eqquasidisc}
U_{\delta t}\ket{\sigma_s}\ket{0}=e^{-\frac{\gamma\delta t}{2}}\ket{\sigma_s}\ket{0}+\sum_{xs'}\sqrt{\left(1-e^{-\gamma \delta t}\right)T^x_{s's}}e^{i\varphi_{xs}}\ket{\sigma_{s'}}\ket{x}.
\end{equation}
Essentially, this evolution preserves the memory state if no event (i.e., end of timestep) occurs, and writes the event symbol to the ancilla and updates the memory accordingly if it does. It can readily be verified that the memory states here have the same overlaps -- and hence are equivalent -- to those prescribed in the discrete evolution Eq.~\eqref{eqdisc} for any $\delta t$. Note that the value of $\gamma$ is irrelevant, other than controlling the rate at which events occur.

Thus, we can now apply our embedding from the previous section. Notably, we can see that $K_0=\exp(-\gamma\delta t/2)\mathbb{I}$ -- where $\mathbb{I}$ is the identity matrix on the memory space -- and thus $H_{\mathrm{eff}}=(-i\gamma/2)\mathbb{I}$. This corresponds to a trivial evolution between jumps, where the system does not change state when events do not occur. Correspondingly, the embeddings of discrete-time stochastic processes correspond to jump-only trajectories. This is again consistent with the dynamics of the discrete-time evolution of quantum simulators of such processes, where the memory undergoes transitions between a discrete set of memory states on each event. The jump operators also follow from the discrete-time evolution, by taking the Kraus operators from Eq.~\eqref{eqdisc} and rescaling by $\sqrt{\gamma}$. Note that this is equivalent to applying Eq.~\eqref{eqjump} to the Kraus operators prescribed by Eq.~\eqref{eqquasidisc}.

\begin{figure}
\includegraphics[width=\linewidth]{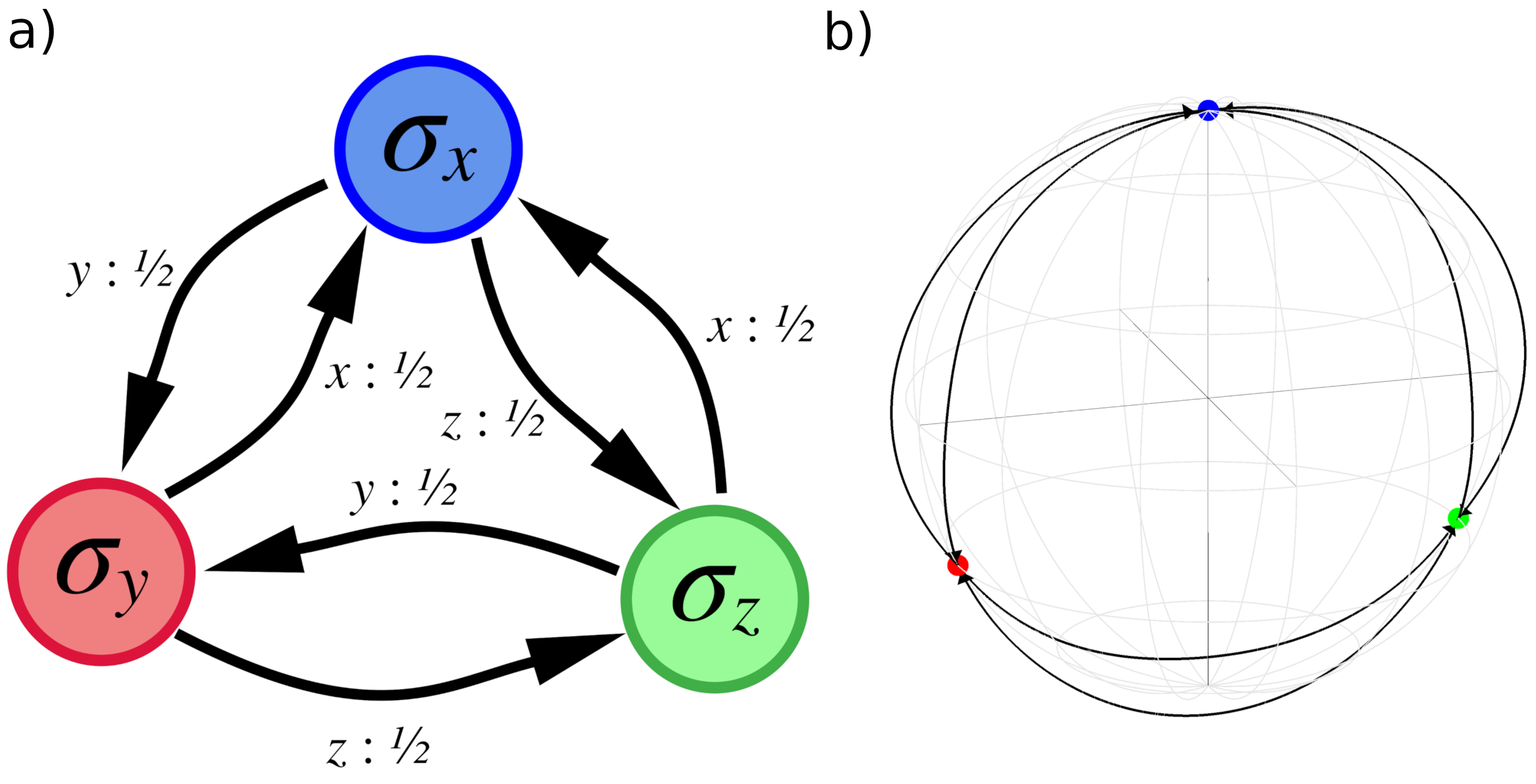}
\caption{(a) HMM representation of the example three state Markov chain described in the main text. (b) Bloch sphere representation of the quantum memory states, showing the location of the three memory states, and black arrows depicting the transitions manifest by the pure jump dynamics of the system.}
\label{fighmmex}
\end{figure}

We illustrate this with an example three-state Markov chain. There are three events $\{x,y,z\}$, and three corresponding states $\{\sigma_x,\sigma_y,\sigma_z\}$. They obey the transition structure $T_{w'w}=0$ if $w=w'$ and $1/2$ otherwise, $\forall w,w'\in\{x,y,z\}$. That is, the system never repeats the same event on two consecutive timesteps, and instead exhibits one of the other two events with equal probability. It has previously been shown that such a process can be modelled with a quantum simulator with only a single qubit of memory~\cite{liu2019optimal}, using the following evolution to define the memory states:
\begin{align}
U\ket{\sigma_x}\ket{0}&=\frac{1}{\sqrt{2}}(\ket{\sigma_y}\ket{y}+\ket{\sigma_z}\ket{z})\nonumber\\
U\ket{\sigma_y}\ket{0}&=\frac{1}{\sqrt{2}}(\ket{\sigma_x}\ket{x}+\ket{\sigma_z}\ket{z})\nonumber\\
U\ket{\sigma_z}\ket{0}&=\frac{1}{\sqrt{2}}(\ket{\sigma_x}\ket{x}-\ket{\sigma_y}\ket{y}),
\end{align}
where it can be seen that $\ket{\sigma_z}=\ket{\sigma_y}-\ket{\sigma_x}$. Without loss of generality, we can assign $\ket{\sigma_x}=\ket{0}$, and subsequently, $\ket{\sigma_y}=(1/2)(\ket{0}+\sqrt{3}\ket{1})$, from which it follows that $\ket{\sigma_z}=(1/2)(-\ket{0}+\sqrt{3}\ket{1})$. From this, we obtain:
\begin{equation}
H_{\mathrm{eff}}=\begin{pmatrix}-\frac{i\gamma}{2} & 0\\ 0 & -\frac{i\gamma}{2}\end{pmatrix},\quad J_x=\sqrt{\gamma}\begin{pmatrix}0 & \frac{\sqrt{2}}{\sqrt{3}}\\ 0 & 0\end{pmatrix},\quad J_y=\frac{\sqrt{\gamma}}{2\sqrt{2}}\begin{pmatrix}1 & -\frac{1}{\sqrt{3}}\\ \sqrt{3} & -1\end{pmatrix},\quad J_z=\frac{\sqrt{\gamma}}{2\sqrt{2}}\begin{pmatrix}-1 & -\frac{1}{\sqrt{3}}\\\sqrt{3} & 1\end{pmatrix}.
\end{equation}
As with the previous example, we plot the corresponding trajectories in \figref{fighmmex}, where it can be seen that it consists solely of jumps between the three memory states.

\section{Classifying structural complexity in open quantum systems?}
\label{secrev}

This embedding of quantum stochastic models as quantum trajectories presents a further enticing opportunity when viewed from the opposite perspective. That is, can we find the model corresponding to observable behaviour of an open quantum system? This would provide a means to apply the full framework of computational mechanics to quantum processes, allowing for the study of the structure of such processes, and the `intrinsic computation'~\cite{crutchfield1994calculi} realised by their dynamics.

However, the contraposition of low-dimensional quantum systems being able to replicate the behaviour of many complex classical stochastic processes is, simply put, that even low-dimensional quantum systems often give rise to statistics that classically appear highly complex~\cite{elliott2020extreme}. Indeed, a pure state of a $D$-dimensional quantum system is described by $2(D-1)$ continuous real parameters. With an appropriate adaptive monitoring scheme, it is possible to pin the system to remain with a finite number of states -- though determining the smallest such `physically-realisable ensemble' remains an interesting open question~\cite{karasik2011many, warszawski2019open}. In general however, the trajectory of an open quantum system may take it through all possible states in its state space. 

Each of these states will in general give rise to different observable future statistics, and thus each possible assignment of the $2(D-1)$ parameters would correspond to a different causal state. This makes it a far from trivial task to construct classical models -- such as the $\varepsilon$-machine -- of the observable behaviour of open quantum systems. A promising way to tackle this is to use a suite of recently-developed tools for analysing structural complexity in classical processes with an uncountably-infinite number of causal states~\cite{jurgens2021divergent}. This is beyond the scope of the present work.

Nevertheless, we can make some headway under certain assumptions on the dynamics of the system. Particularly, let us assume that the jump operators are erasing, in the sense that each jump operator maps all states into the same state. That is, $J_x=\sum_{j}a_j^x\ket{\psi_x}\bra{j}$ for some $\{a_j^x\}$ and $\ket{\psi_x}$ for all $x$. Then, we need only one continuous parameter (in addition to the discrete parameter corresponding to the last jump label) to describe the state of the system at all times. That is, if the last jump to occur was $J_x$, and a time $t$ has elapsed since then, then the (non-normalised) system state is given by $\exp(-iH_{\mathrm{eff}}t)\ket{\psi_x}$.

The observable behaviour then takes the form of a semi-Markov process, where the symbolic component of the dynamics can be expressed as a Markov chain, but the times between events are stochastic variables that depend on the last event. This is a special case of the HSMMs introduced above, where the modes correspond to the most recent event. Let us define $\ket{\psi_x(t)}:=\exp(-iH_{\mathrm{eff}}t)\ket{\psi_x}$. The modal survival probabilities then take the form $\Phi_x(t)=\braket{\psi_x(t)}{\psi_x(t)}$. From the infinitesimal evolution, we can also deduce that the probability that event $x'$ occurs in the interval $[t,t+\delta t)$ given last event $x$ is $\bopk{\psi_x(t)}{J_{x'}^\dagger J_{x'}}{\psi_x(t)}\delta t$. Putting this together we have
\begin{align}
\label{eqsimpleconverse}
P(x',t|x)&=\bopk{\psi_x}{e^{iH_{\mathrm{eff}}^\dagger t}J_{x'}^\dagger J_{x'}e^{-iH_{\mathrm{eff}} t}}{\psi_x},\nonumber\\
T_{x'x}&=\int_0^\infty P(x',t|x) dt,\nonumber\\
\phi_{x'x}(t)&=P(x',t|x)/T_{x'x}.
\end{align}
With this description, the standard tools of computational mechanics can be applied to investigate the structure of the process. We leave such a dissection of physically-relevant quantum processes for future work.

More generally, for jumps that are not erasing the state of the system will typically depend on the entire history of jumps and jump times, i.e., $\past{\bm{x}}$. We can define this associated state as an encoded memory state $f_q(\past{\bm{x}})$, and the distributions must now condition upon $f_q(\past{\bm{x}})$ rather than the previous symbol $x$ alone. Such a distribution meaningfully exists for all accessible $f_q(\past{\bm{x}})$, and will generically be distinct for $f_q(\past{\bm{x}})\neq f_q(\past{\bm{x}}')$; the causal states are then associated with equivalence of the iterated form of these distributions (i.e., the $P(\fut{\bm{X}}|f_q(\past{\bm{X}})$). As remarked above however, this will in general yield an infinite number of causal states.

Note however, that the standard method for simulating quantum trajectories is implicitly based on the construction of such distributions~\cite{daley2014quantum}. Recall that given a post-jump state $\ket{\psi}$, the time of the next jump is determined by the time at which $\bopk{\psi}{e^{iH_{\mathrm{eff}}^\dagger t}e^{-iH_{\mathrm{eff}} t}}{\psi}<r$ for some randomly generated $r\in[0,1]$. This quantity corresponds to the survival probability $\Phi_{\ket{\psi}}(t)$of the initial state $\ket{\psi}$ under the effective non-Hermitian Hamiltonian $H_{\mathrm{eff}}$, which equivalently corresponds to $1-\sum_x\int_0^tP(x,t'|\ket{\psi})dt'$. The specific jump is then determined by sampling from the distribution $P(x|t,\ket{\psi})=P(x,t|\ket{\psi})/\sum_x P(x,t|\ket{\psi})$. Yet unlike our need to construct all possible such distributions to meaningfully apply the framework of computational mechanics, the simulation of quantum trajectories is much less demanding. Such simulation requires only that we consider the distributions associated with the post-jump states visited on the trajectory, and even then, the monotonicity of $\Phi_{\ket{\psi}}(t)$ can be used to circumvent the need to construct the full distribution to determine the point at which it coincides with $r$.

\section{Discussion} 
\label{secdis}

In this work we have established a means by which memory-efficient quantum simulators of stochastic processes can be embedded within the natural evolution of monitored open quantum systems. The monitored dissipation of the system corresponds to the observed events in the process, and so each quantum trajectory charting a particular monitored evolution of the system corresponds to a realisation of the stochastic process by the embedded quantum model. This resolves a gap between the continous-time nature of the processes modelled, and the quasi-continuous nature of the model evolution itself.

While seemingly innocuous, the resolution of this gap has important ramifications. Foremost, it removes the need for a (semi-)arbitrary timestep size in the model with the added benefit of removing the need for an external control to implement the evolution at each timestep. Crucially, this makes the model autonomous, and not reliant on an external timekeeping device to synchronise timestep sizes to. There are also potential practical considerations of benefit. Implementing a quasi-continuous model requires an ever-growing number of gates as the timestep size decreases (with ever increasing precision required of each step), also leading to an ever-increasing degree of susceptibility to noise in the implementation. By embedding directly as a continuous evolution of a quantum system we circumvent this, and instead our sources of error instead come down to how well we can instantiate the appropriate natural Hamiltonian and jump operators. These do not scale with the precision of our model -- and indeed, together with the timescale on which we can resolve jumps, they can be seen to implicitly define the meaningful precision that can be achieved. Interestingly, this moves counter to the norm in scientific computing; rather than converting a problem into a digital computation to solve it, we are instead embedding a digital computation within the natural evolution of a system. This is, in essence, a form of analogue quantum simulation for classical stochastic dynamics.

There are a number of natural extensions. The embedding itself can be extended into the regime of input-output processes -- where the behaviour of the system can be influenced by stimuli from its environment~\cite{elliott2022quantum}. This will enable the realisation of quantum models of adaptive agents in true continuous-time. Moreover, such an embedding for the input-output domain may prove fruitful in probing the structure of general quantum stochastic processes with intervention~\cite{milz2021quantum} in continuous-time. We have also begun laying down the framework for using the embedding to apply tools from complexity science to understand structure in open quantum systems. Indeed, the application of such tools to many-body quantum states has already yielded interesting early results, such as the correspondence of sharp peaks in certain measures of complexity with quantum phase transitions~\cite{suen2022surveying}. Applying these ideas to the dynamics of quantum systems may provide fascinating insights into the structure of non-equilibrium quantum steady-states~\cite{prosen2009matrix}, measurement-induced phase transitions~\cite{skinner2019measurement}, and quantum chaos~\cite{haake1991quantum}. Finally, by demonstrating that our quantum models of continuous-time stochastic processes can indeed be realised in an autonomous, continuous-time manner, we have affirmed that they do indeed provide a viable means of implementing autonomous quantum clocks~\cite{woods2021autonomous}. This connection may yield profitable means of applying results from quantum stochastic simulation to gain a deeper understanding of quantum clocks (and vice versa), including the fundamental resources needed to track time.

\appendix

\section{Convergence of memory state overlaps}

Recall that the memory states of the quantum models are implicitly defined by the evolution equation Eq.~\eqref{eqexactct}:
\begin{equation}
U_{\delta t}\ket{\varsigma_{gt}}_{\delta t}\ket{0}=\sqrt{\frac{\Phi_g(t+\delta t)}{\Phi_g(t)}}\ket{\varsigma_{gt+\delta t}}_{\delta t}\ket{0}
+\sum_{xg'}\sqrt{\frac{\int_t^{t+\delta t}T_{g'g}^x\phi_{g'g}^x(t')dt'}{\Phi_g(t)}}\ket{\varsigma_{g'0}}_{\delta t}{\ket{x}}.
\end{equation}
Correspondingly, using that $U^\dagger_{\delta t}U_{\delta t}=\mathbb{I}$, we have that the state overlaps are given by
\begin{align}
\braket{\varsigma_{gt}}{\varsigma_{g't'}}_{\delta t}=&\sqrt{\frac{\Phi_g(t+\delta t)\Phi_{g'}(t+\delta t)}{\Phi_g(t)\Phi_{g'}(t')}}\braket{\varsigma_{gt+\delta t}}{\varsigma_{g't'+\delta t}}_{\delta t}\nonumber\\ &+\sum_x\sum_{g''g'''}\sqrt{\frac{T^x_{g''g}T^x_{g'''g'}\int_t^{t+\delta t}\int_t^{t'+\delta t}\phi_{g''g}^x(t'')\phi_{g'''g'}^x(t'')dt''dt'''}{\Phi_g(t)\Phi_{g'}(t')}}\braket{\varsigma_{g''0}}{\varsigma_{g'''0}}_{\delta t}.
\end{align}
Recall that the mode into which the system transitions is a deterministic function of the current mode and event. Let us denote this by the function $\lambda(g,x)$. Then, we have that
\begin{equation}
\braket{\varsigma_{gt}}{\varsigma_{gt+\delta t}}_{\delta t}=\sqrt{\frac{\Phi_g(t+2\delta t)}{\Phi_g(t)}}\braket{\varsigma_{gt+\delta t}}{\varsigma_{gt+2\delta t}}_{\delta t}+\sum_xT^x_{\lambda(g,x)g}\sqrt{\frac{\int_t^{t+\delta t}\int_{t+\delta t}^{t+2\delta t}\phi_{\lambda(g,x)g}^x(t')\phi_{\lambda(g,x)g}^x(t'')dt'dt''}{\Phi_g(t)\Phi_g(t+\delta t)}}.
\end{equation}
By repeatedly iterating through the first term on the right-hand side, we then obtain
\begin{equation}
\label{eqappoverlap}
\braket{\varsigma_{gt}}{\varsigma_{gt+\delta t}}_{\delta t}=\sum_xT^x_{\lambda(g,x)g}\sum_{n=0}^\infty\sqrt{\frac{\int_{t+n\delta t}^{t+(n+1)\delta t}\int_{t+(n+1)\delta t}^{t+(n+2)\delta t}\phi_{\lambda(g,x)g}^x(t')\phi_{\lambda(g,x)g}^x(t'')dt'dt''}{\Phi_g(t)\Phi_g(t+\delta t)}}.
\end{equation}

Let us now consider the case where the $\phi_{g'g}^x(t)$ are all everywhere continuous, i.e., that for all $\epsilon>0$, there exists a $\delta t(\epsilon)$ such that $|t-t'|<\delta t$ implies $|\phi_{g'g}^x(t)-\phi_{g'g}^x(t')|<\epsilon$ for all $t,t'$. It follows that there then exists a $\delta t(\epsilon)$ such that $|t-t'|<\delta t$ implies $|\int_t^{t+\Delta t}\phi_{g'g}^x(\tau)d\tau-\int_{t'}^{t'+\Delta t}\phi_{g'g}^x(\tau)d\tau|<\epsilon\Delta t$. Since $\epsilon$ can be made arbitrarily small, it then follows that the correction to Eq.~\eqref{eqappoverlap} by replacing $\sqrt{\int_{t+n\delta t}^{t+(n+1)\delta t}\int_{t+(n+1)\delta t}^{t+(n+2)\delta t}\phi_{\lambda(g,x)g}^x(t')\phi_{\lambda(g,x)g}^x(t'')dt'dt''}$ with $\int_{t+n\delta t}^{t+(n+1)\delta t}\phi_{\lambda(g,x)g}^x(t')dt'$ can be made arbitrarily small. 

Relaxing such that the $\phi_{g'g}^x(t)$ are almost everywhere continuous, and everywhere finite, the corrections to the above are of finite magnitude and have zero measure. Recalling that $\Phi_g(t):=\sum_{xg'}T_{g'g}^x\int_t^\infty\phi_{g'g}^x(t')dt'$, we have that for sufficiently small $\delta t$, $\Phi_g(t)-\Phi_g(t+\delta t)\propto\delta t$. Putting this all together, we obtain that at sufficiently small $\delta t$, $1-\braket{\varsigma_{gt}}{\varsigma_{gt+\delta t}}_{\delta t}\propto\delta t$, i.e., the infinitesimal evolution leads to an infinitesimal change in the memory state.

\section{Details of continuous-time example}

Recall that in our example we have two Poissonian decay channels with rates $\gamma_1$ and $\gamma_2$, leading to the emission of symbols 1 and 2 respectively. After decay event $x$ the system will transition into mode $g_x$, where channel $x$ is chosen with probability $p$, and the other channel with probability $\bar{p}=1-p$. That is, $T^x_{g_{x'}g_{x''}}$ equals $p$ if $x=x'=x''$, $\bar{p}$ if $x=x'\neq x''$, and zero otherwise. The dwell time distributions take the form $\phi^x_{g_{x'}g_{x''}}=\gamma_x\exp(-\gamma_xt)$.

A viable choice of memory states for a causal model is to assign each pair $(g,t)$ to a distinct memory state~\cite{marzen2017structure}. The steady-state probability of these memory states is given by $P((g,t))=\mu\pi_g\Phi_g(t)$, where $\pi_g$ is the probability that the system is in mode $g$ immediately after emission (which can be calculated from the fixed point of $\sum_xT^x_{g'g}$), and $\mu^{-1}:=\sum_{gg'x}\pi_g\int_0^\infty tT^x_{g'g}\phi_{g'g}^x(t)dt$~\cite{elliott2019memory}. 

However, this is not minimal, as the conditional distribution describing what the next event is -- and when it will occur -- coincides (with an offset) for the two modes. Consider, the conditional distributions, given by $P(x,\tfut|g_1,\tpast)\propto p\gamma_1\exp(-\gamma_1(\tpast+\tfut))+\bar{p}\gamma_2\exp(-\gamma_2(\tpast+\tfut))$ and $P(x,\tfut|g_2,\tpast)\propto \bar{p}\gamma_1\exp(-\gamma_1(\tpast+\tfut))+p\gamma_2\exp(-\gamma_2(\tpast+\tfut))$. We can see that over time, each distribution becomes increasing weighted in favour of the channel with the slower decay rate. Without loss of generality let this be channel 2, such that $\gamma_1>\gamma_2$. Then, we can see the two distributions coincide with an offset $\tau$ such that $\exp((\gamma_1-\gamma_2)\tau)=p^2/\bar{p}^2$ for $p>\bar{p}$, or $\exp((\gamma_1-\gamma_2)\tau)=\bar{p}^2/p^2$ for $p<\bar{p}$. The causal states then correspond to the merging of the memory states $(g,t)$ according to this offset equivalence, e.g., for $p>\bar{p}$ and $\gamma_1>\gamma_2$, $(g_1,t+\tau)\sim_\varepsilon(g_2,t)$.  Nevertheless, the continuous nature of the causal states ensures that there is an infinite number of such states, and hence the memory cost of the $\varepsilon$-machine diverges.

Interestingly, the procedure by which we construct our quantum models takes care of this merging of states automatically~\cite{elliott2019memory}. To construct the quantum model we define the following pair of `generator' states $\ket{\varphi_x}$ that satisfy~\cite{elliott2021quantum} 
\begin{equation}
U_{\delta t}\ket{\varphi_x}\ket{0}=e^{-\gamma_x\frac{\delta t}{2}}\ket{\varphi_x}\ket{0}+\sqrt{1-e^{-\gamma_x\delta t}}\ket{\sigma_x0}_{\delta t}\ket{x}.
\end{equation}
From $\braket{\varphi_x}{\varphi_{x'}}=\bopk{\varphi_x}{U_{\delta t}^\dagger U_{\delta t}}{\varphi_{x'}}$ it then follows that $\braket{\varphi_x}{\varphi_{x'}}=\delta_{xx'}$, and thus we can without loss of generality assign $\ket{\varphi_1}=\ket{0}$ and $\ket{\varphi_2}=\ket{1}$.

It can then be seen from direct substitution that for this example process the quasi-continuous evolution operator equation Eq.~\eqref{eqexactct} is satisfied by setting
\begin{align}
\ket{\sigma_{1t}}_{\delta t}&=\frac{\sqrt{p}e^{-\gamma_1\frac{\delta t}{2}}\ket{0}+\sqrt{\bar{p}}e^{-\gamma_2\frac{\delta t}{2}}\ket{1}}{\sqrt{\Phi_1(t)}}\nonumber\\
\ket{\sigma_{2t}}_{\delta t}&=\frac{\sqrt{\bar{p}}e^{-\gamma_1\frac{\delta t}{2}}\ket{0}+\sqrt{p}e^{-\gamma_2\frac{\delta t}{2}}\ket{1}}{\sqrt{\Phi_2(t)}}.
\end{align}
It can be verified that these states coincide with the appropriate offset as described above, and hence quantum memory states belonging to the same causal state are identical. This merging can also be seen in \figref{fighsmmex}(b). These quantum memory states can then be used to determine $U_{\delta t}$, and consequently, the associated Kraus operators. 

We obtain that 
\begin{equation}
K_{\delta t}^0=\begin{pmatrix} e^{-\gamma_1\frac{\delta t}{2}} & 0\\ 0& e^{-\gamma_2\frac{\delta t}{2}}\end{pmatrix},
\end{equation}
and hence from Eq.~\eqref{eqheffembed} deduce that
\begin{equation}
H_{\mathrm{eff}}=\begin{pmatrix}-i\frac{\gamma_1}{2} & 0\\ 0 & -i\frac{\gamma_2}{2}\end{pmatrix}.
\end{equation}
We can also readily obtain the associated jump operators. We have that $J_x=\sqrt{\gamma_x}\ket{\sigma_{x0}}\bra{0}$, and hence
\begin{equation}
J_1=\sqrt{\gamma_1}\begin{pmatrix}\sqrt{p}\\\sqrt{\bar{p}}\end{pmatrix}\begin{pmatrix} 1 & 0 \end{pmatrix}=\begin{pmatrix}\sqrt{\gamma_1p} & 0 \\ \sqrt{\gamma_1\bar{p}} & 0 \end{pmatrix}.
\end{equation}
Similarly,
\begin{equation}
J_2=\sqrt{\gamma_2}\begin{pmatrix}\sqrt{\bar{p}}\\\sqrt{p}\end{pmatrix}\begin{pmatrix} 0 & 1 \end{pmatrix}=\begin{pmatrix} 0 & \sqrt{\gamma_1\bar{p}} \\ 0 & \sqrt{\gamma_1p} \end{pmatrix}.
\end{equation}

\acknowledgments
This work was funded by the University of Manchester Dame Kathleen Ollerenshaw Fellowship, the Imperial College Borland Fellowship in Mathematics, the Lee Kuan Yew Endowment Fund (Postdoctoral Fellowship), grants FQXi-RFP-1809 and FQXi-RFP-IPW-1903 from the Foundational Questions Institute and Fetzer Franklin Fund (a donor advised fund of Silicon Valley Community Foundation), the National Research Foundation, Singapore, and Agency for Science, Technology and Research (A*STAR) under its QEP2.0 programme (NRF2021-QEP2-02-P06), and the Singapore Ministry of Education Tier 1 Grants RG190/17 and RG77/22 and Tier 2 Grant MOE-T2EP50221-0005. T.J.E.~thanks the Centre for Quantum Technologies for their hospitality.

\bibliography{ref}

\end{document}